\documentstyle[aps,preprint]{revtex}
\draft

\begin{document}
\widetext

\title{Instabilities in a Two-Component, Species Conserving Condensate}
\author{
A. G. Rojo} \address{  Department of Physics, 
The University of Michigan, Ann Arbor, Michigan
48109-1120, USA} 
\maketitle

\widetext
\begin{abstract}

We consider a  system of two species of bosons
of equal mass, with 
interactions $U^{a}(|{\mathbf{x}}|)$
 and $U^{x}(|{\mathbf{x}}|)$ for bosons
of the same and different species respectively.
We present a rigorous proof--valid when the Hamiltonian does not include a species switching term--showing that, when 
$U^{x}(|{\mathbf{x}}|)>U^{a}(|{\mathbf{x}}|)$, the ground state is fully ``polarized"
(consists of atoms of one kind only). In the unpolarized phase the low energy
excitation spectrum corresponds to two linearly dispersing modes that are even and 
odd under species exchange. The polarization instability is signaled by the vanishing
of the velocity of the odd modes.

\pacs{PACS numbers: 03.75Fi, 05.30.-d}
\end{abstract}

The experimental observation of Bose--Einstein condensation (BEC) in dilute atomic 
systems\cite{exp_bec}
has triggered a very intense theoretical activity\cite{rmp}. Attention has
broadened to include
  condensates with internal degrees of freedom, or multi--species BEC, which were  realized 
  for trapped rubidium\cite{myatt} and sodium\cite{stamper}.
  Early theoretical analysis of $m$--species condensates focuses on the
  large $m$ limit for Hamiltonians invariant under $U(m)$ 
  transformations\cite{halperin}, superfluid helim mixtures\cite{khalatinov}
  and spin--polarized hydrogen\cite{siggia}.
  In the context of BEC much of the theoretical attention concentrates on
  spinor condensates\cite{spinor} in which the internal degrees of freedom 
  correspond to the different Zeeman states of the of 
  a particular hyperfine manifold such as the $f=1$ manifold in sodium. 
  In these cases the two--body interaction is invariant under under rotation
  in species space. On the other hand, for rubidium one has 
  two non--degenerate internal states corresponding to two different manyfolds,
   and the two--body interaction is not 
  invariant under $SU(2)$ rotations in species space\cite{esry}.
The ground state and excitation spectrum of a two--species  ($a$ and $b$) condensate
 of this 
kind was studied in mean field in Ref.\cite{meystre}, where it was shown that the quasiparticle
energy can become imaginary, signaling an instability
 when the inter--species repulsion $U^{x}$ is larger than the intra--species repulsion
  $U^{a}$. This kind of  treatment follows--as do the majority of theoretical approaches to
  BEC--the Gross--Pitaevskii (GP) mean field  equation\cite{gross}.
  On the other hand, rigorous results  for BEC (and many body problems in general) are
   scarce and at the same time 
   useful in providing control for approximate solutions.  
 With this motivation, in this paper we consider a two--species system of interacting bosons,
and show rigorously that the instability  mentioned above corresponds to the tendency of the system
to  ``polarize", the true ground state consisting of only one species of bosons 
when $U^{x}>U^{a}$. We also discuss the low--energy excitation spectrum and show
 that the instability is signaled by a divergence of the ``compressibility" associated
 with exchanging particles from one species to the other at fixed total particle number.

The Hamiltonian for our two-species system ($a$ and $b$) of bosons of equal mass
$m$ is given by ($\hbar =1$)
\begin{equation}
H=K+U_{a}+U_{b}+U_{ab},\end{equation}
with
\begin{eqnarray}
K&=&\sum_{\mathbf{k}}{ k^2\over 2m}\left[
\psi_a^{\dagger }({\mathbf{k}})%
\psi_a^{}({\mathbf{k}})+
\psi_b^{\dagger }({\mathbf{k}})%
\psi_b^{}({\mathbf{k}})
\right] 
\\
U_{a}+U_{b}&=&\frac{1}{2}
\int d^3x\,d^3y\, U^a(|{\bf x}-{\bf y}|) \left[ \rho_a({\bf x})\rho_a({\bf y})
+\rho_b({\bf x})\rho_b({\bf y})
\right]
\\U_{ab} &=&\int d^3x\,d^3y\, U^x(|{\bf x}-{\bf y}|)  \rho_a({\bf x})\rho_b({\bf y}).
  \label{h1}
\end{eqnarray}

In the  above equations $K$ is the kinetic term, and the  terms
$(U_{a}+U_{b})$ and  $U_{ab}$ correspond
 to the interaction between bosons of the same  and different 
 species respectively. The operators $\psi_i^{}({\mathbf{x}})$ 
 destroy a boson of species $i$ ($i=a,b$) at position ${\mathbf{x}}$, and
 obey the following commutation relations $[\psi_i^{\dagger}({\mathbf{x}}),
 \psi_j^{}({\mathbf{x'}})]=\delta_{i,j}\delta ({\mathbf{x}}-{\mathbf{x}'})$.
 Also, $\rho_i({\bf x})=\psi^{\dagger}_i({\bf x})\psi_i({\bf x})$, and 
  $\psi_i^{}({\mathbf{k}})=V^{-1/2}\int d^3x \, \exp(i{\bf k}\cdot{\bf x}) 
\psi_i^{}({\mathbf{x}})$ with $V$ the total volume.

Using a variational argument we will prove that, for  potentials satisfying
$U^x(|{\mathbf{x}}|)>U^a(|{\mathbf{x}}|)$ the ground state of the above Hamiltonian is completely polarized.
By ``polarized" we mean that either of the two situations is realized:
$\left\langle \hat{N}_{a}\right\rangle =0,\left\langle \hat{N}_{b}\right\rangle =N$; or $%
\left\langle \hat{N}_{a}\right\rangle =N,\left\langle \hat{N}_{b}\right\rangle =0,$
with $N$ the total particle number
and  $\hat{N}_{i}=\int d^3x \,  \rho_i({{\bf{x}}})$.

We note that, since the masses are the same for both species, 
the kinetic term of  Hamiltonian (\ref{h1}) commutes with the
``rising'' operator defined as:
\begin{equation}
{\cal{O}}_{R}=\int d^3x \, 
\psi_a^{\dagger}({\mathbf{x}})
\psi_b^{}({\mathbf{x}})
\equiv \sum_{\bf k} 
\psi_a^{\dagger}({\mathbf{k}})\psi_b^{}({\mathbf{k}})
\end{equation}
 which conserves the total number of particles but converts
particles of type $b$ into particles of type $a.$ More specifically,
for the case of different masses for each species we have that
$[{\cal{O}}_{R},K]=\left({1\over 2m_b}-{1\over 2m_a}\right)\sum_{\bf k}k^2 
\psi_a^{\dagger}({\mathbf{k}})\psi_b^{}({\mathbf{k}})$.

We have in mind the alkali atoms, which have a hard core interaction, meaning that
the exact wave functions vanish when the coordinates of two atoms of either 
species coincide.

 Since the Hamiltonian conserves
the particle number for each species, we can start with
 the normalized ground state wave function $|\Psi_0(N_a,N_b)\rangle$ in the 
subspace of $N_a$ ($N_b$) particles of species $a$($b$). 
The unpolarized situation corresponds to $N_a=N_b=N/2$.
Now let us consider the 
normalized variational wave function $|\Psi_v(N_a+N_b,0)\rangle$ 
obtained by the action of the rising operator on 
$|\Psi_0(N_a,N_b)\rangle$ $N_b$ times:
\begin{equation}
|\Psi_v(N_a+N_b,0)\rangle={1\over\sqrt{N_b!}}({\cal{O}_{R}})^{N_b}|\Psi_0(N_a,N_b)\rangle.
\label{psiv}
\end{equation}

Note that $\Psi_v$ represents a completely polarized wave function, with particles
 of species $a$ only. Since $[{\cal{O}_{R}},K]=0$, we have
 
 \begin{equation}
\langle \Psi_v(N_a+N_b,0)|K |\Psi_v(N_a+N_b,0)\rangle
=\langle \Psi_0(N_a,N_b)|K |\Psi_0(N_a,N_b)\rangle,
 \end{equation}
 meaning that the completely polarized variational wave function and the ground state 
 of the subspace $(N_a,N_b)$
 have the same expectation value of the 
 kinetic energy. We stress that the function $|\Psi_v(N_a+N_b,0)\rangle$ as defined
 in Eq. (\ref{psiv}) is normalized only because the exact wave function 
 $|\Psi_0(N_a,N_b)\rangle$ vanishes when any two coordinates coincide. Otherwise we
 would have to worry about permutation factors whenever coordinates coincide.

In order to compute the change in the potential
 energy we write the expectation values of $U^a$ and $U^x$ as an integral
 over all the multiparticle configurations $\Gamma\equiv(\Gamma_a,\Gamma_b)$
 with coordinates  $\{{\bf X}\}_\Gamma\equiv \{{\bf{x}}_{1,\Gamma_a},\cdots, {\bf{x}}_{N_a,\Gamma_a}
 ;{\bf{x}}_{1,\Gamma_b},\cdots, {\bf{x}}_{N_b,\Gamma_b}\}$. 
 For each configuration $\Gamma$ let us regard the particle coordinates 
 $\{{\bf X}\}_\Gamma$
 as nodes of a graph. There are $N_a(N_a-1)/2$ and $N_b(N_b-1)/2$ edges connecting 
 pairs of particles of species $a$ and $b$ respectively,  and 
 $N_aN_b$ edges connecting a particles of different species.
 The contribution to the expectation value of the potential 
 energy $U_0=\langle U_{a}\rangle
 +\langle U_{b}\rangle+\langle U_{ab}\rangle$ from this configuration
  is a function of the length of the edges of the graph, which 
 can be classified in three sets:
 $\{ {\bf{\ell}}^{a}_{1,\Gamma_a},\cdots, {\bf{\ell}}^{a}_{N_a(N_a-1)/2,\Gamma_a}\}$,
 $\{ {\bf{\ell}}^{b}_{1,\Gamma_b},\cdots, {\bf{\ell}}^{b}_{N_b(N_b-1)/2,\Gamma_b}\}$,
 and
 $\{ {\bf{\ell}}^{ab}_{1,\Gamma},\cdots, {\bf{\ell}}^{ab}_{N_aN_b,\Gamma}\}
 $, where $ {\bf{\ell}}^{a}_{i,\Gamma_a}$ is one of the possible 
 lenghts $|{\bf{x}}_{k,\Gamma_a}-{\bf{x}}_{l,\Gamma_a}| $, etc. 
 The potential energy is therefore given by
 
 \begin{equation}
  U_0=\int d^{3N} X_\Gamma \,|\Psi_0({\bf X}_\Gamma)|^2 \left[
  \sum_{i=1}^{{N_a(N_a-1)/2}}U^{a}({{\bf{\ell}}^{a}_{i,\Gamma_a}})
  +
  \sum_{i=1}^{{N_b(N_b-1)/2}}U^{a}({{\bf{\ell}}^{b}_{i,\Gamma_b}})
  +
  \sum_{i=1}^{{N_aN_b}}U^{x}({{\bf{\ell}}^{ab}_{i,\Gamma}})
  \right],
  \end{equation}
with $\Psi_0({\bf X}_\Gamma)$ the ground state wave--function in first quantization.

In the variational wave function all the edges of type $b$  and $ab$ 
are both changed to edges of type $a$. Therefore  the contribution to the
potential energy of each edge of configuration
$\Gamma$ changes according to $U^{a}({{\bf{\ell}}^{b}_{i,\Gamma_b}})\rightarrow
U^{a}({{\bf{\ell}}^{b}_{i,\Gamma_b}})$,
 $U^{x}({{\bf{\ell}}^{ab}_{i,\Gamma}})\rightarrow
U^{a}({{\bf{\ell}}^{ab}_{i,\Gamma}})$.
If we call $U_v$ the expectation value of the potential energy in the variational wave
function, and $E_v$ the variational wave function, we obtain that $\Delta E=E_v -E_0=U_v-U_0$
 is given by
 \begin{equation}
\Delta E=
\int d^N X_\Gamma \,|\Psi_0({\bf X}_\Gamma)|^2
\sum_{i=1}^{N_aN_b} \left[ U^{a}({{\bf{\ell}}^{ab}_{i,\Gamma}})
-U^{x}({{\bf{\ell}}^{ab}_{i,\Gamma}})\right].
\end{equation}

We see that when the 
interactions satisfy $U^x(|{\mathbf{x}}|) > U^a(|{\mathbf{x}}|)$ 
for all values of ${\bf{x}}$, 
 $\Delta E$ is a sum of negative terms. 
 In general the potentials have a repulsive short-range term and a long-range
 attractive tail.
The condition for validity of our proof  is that
there are no "crossings" of the potentials  $U^x(|{\mathbf{x}}|)$ 
and $U^a(|{\mathbf{x}}|)$ as a 
function of the coordinate. The simplest approximation will be to take the
attractive components of both potentials as equivalent (of the form $-C_6/R^6$)
and differing short range components.  This dependence is consistent
with calculations for ultra-cold Na collisions\cite{julienne}.
 
 Since the completely polarized variational
wave function has  lower energy than the true ground state of the partially polarized
subspace, the gound state will be completely polarized. 
It is also  evident that the proof is also valid in the case  different 
intra--species interactions $U^a({\mathbf{x}})$ and $U^b({\mathbf{x}})$.
 As long as $U^{x} >
 U^{a},U^{b}$, the ground state is polarized with particles 
of type $a$ ($b$) when $U^a(|{\mathbf{x}}|) < U^b(|{\mathbf{x}}|)$ ($
(U^b(|{\mathbf{x}}|) < U^a(|{\mathbf{x}}|)$). 
This means that the difference $U^b(|{\mathbf{x}}|) - U^a(|{\mathbf{x}}|)$ plays the
 role of a ``symmetry breaking field". Also, note that the above proof is also valid 
 in the presence of an external potential $U_e({\mathbf{x}})$ that is equal for both species,
 meaning that the polarization transition also occurs for trapped atoms.
 
 We note that the bosonic nature of the particles is crucial for establishing our 
 rigorous proof. If the atoms $a$ and $b$ were fermions,  the variational argument
  ceases to be valid: since the Hamiltonian
 conserves species, atoms belonging to different species can be considered distinguishable, 
 and the wave function does not change sign if we exchange any two atoms $a$ and $b$
 following a path $P$. If we convert
 an atom $a$ to an atom $b$, the wave function has to change sign under
 the particle exchange following the same path $P$, 
 implying that we have to introduce an additional node in the wave function.
 Formally this means that  ${\cal{O}}_{R}|\Psi_0(N_a,N_b)\rangle=0$ for fermions.

 We now discuss the low energy excitation spectrum,
 which in the symmetric case  ($U^{a} = U^{b}$) 
 can be computed  using  longitudinal sum rules\cite{huang1,huang2}.
 In the unpolarized case 
 the excitations correspond to two phonon branches with wave functions  $
 \rho ^{\pm}_{\bf k} =
 (\rho ^{a}_{\bf k}\pm\rho ^{b}_{\bf k})|0\rangle$.
 The operators $\rho ^{i}_{\bf k}=V^{-1/2}\int d^3 x e^{i{\bf k}\cdot {\bf x}}\rho_{{\bf x},i}
$ are the Fourier transforms of the density operators for each species. In other words, since the Hamiltonian is invariant 
under exchange of species, the excitations are either even or odd in the species index.
 
   The excitation energies  are given by 
  \begin{equation}
  \omega^{\pm}_{\bf k}= 
  {\langle 0|\rho_{\bf -k}^{\pm}H\rho_{\bf k}^{\pm}|0\rangle
  \over
  \langle 0|\rho_{\bf -k}^{\pm}\rho_{\bf k}^{\pm}|0\rangle},
  \end{equation}
  where we have shifted the origin of energies ($H \rightarrow H-E_0$; 
  $H|0\rangle =0$). 
  We will first show that  $\langle 0|\rho_{\bf -k}^{\pm}H\rho_{\bf k}^{\pm}|0\rangle
  =nk^2/2m,$ with $n=N/V$.
  Consider 
  \begin{eqnarray}
  \langle 0|\rho_{\bf -k}^{a}H\rho_{\bf k}^{b}|0\rangle&=&
  {1\over 2} 
  \langle 0|\left\{[ \rho_{\bf -k}^{a},H]\rho_{\bf k}^{b}  + 
   \rho_{\bf -k}^{a}  [\rho_{\bf k}^{b},H]
  \right\}|0\rangle  \label{ab1} \\
  &=& {\hbar^2\over 2m}\sum_{{\bf q},{\bf q}'}
  {\bf k}\cdot{\bf q}\langle 0|\left[
  \psi_a^{\dagger}({{\bf q}'-{\bf k}})
  \psi_a^{}({{\bf q}'})
  \psi_b^{\dagger}({{\bf q}+{\bf k}})
  \psi_b^{}({{\bf q}}) \right. \nonumber \\
  && \left. 
  -\psi_a^{\dagger}({{\bf q}+{\bf k}})
  \psi_a^{}({{\bf q}})
  \psi_b^{\dagger}({{\bf q}'-{\bf k}})
  \psi_b^{}({{\bf q}}')\right]|0\rangle, 
 \label{ab2}
  \end{eqnarray}
  with Eq. (\ref{ab2}) following from the direct evaluation of the commutators in Eq. (\ref{ab1}).
  Since the ground 
  state is symmetric under species exchange, and since the operators $b$ and $a$ commute,
 Eq. (\ref{ab2}) implies that
  $\langle 0|\rho_{\bf -k}^{a}H\rho_{\bf k}^{b}|0\rangle=
  \langle 0|\rho_{\bf -k}^{b}H\rho_{\bf k}^{a}|0\rangle=0$.
 On the other hand, if in Eq. (\ref{ab2}) we replace $b$ with $a$ we obtain the 
 usual sum rule
 $
 \langle 0|\rho_{\bf -k}^{a}H\rho_{\bf k}^{a}|0\rangle=(\hbar^2 k^2/2m) V^{-1}\langle 0|
 \sum_{\bf q} \psi_a^{\dagger}({{\bf q}}) \psi_a^{}({{\bf q}})
 |0\rangle$, which implies that for the even and odd modes the exact $f$--sum 
  rule applies \cite{nozieres}:
  $\langle 0|\rho_{\bf -k}^{\pm}H\rho_{\bf k}^{\pm}|0\rangle
  =nk^2/2m$, or, which is equivalent: 
  \begin{equation}
  n\int d \omega \, \omega  S^{\pm}(k,\omega)={nk^2\over 2m},
  \label{sr1}
  \end{equation} with 
 
\begin{equation}
 S^{\pm}(k,\omega)=n^{-1}\sum_{\nu} |\langle \nu |\rho_{\bf k}^{\pm}|0\rangle |^2
 \delta (\omega - \omega_\nu)
\end{equation}
the dynamic structure factor of the even and odd modes. 
 The excitation energies are therefore given by
  \begin{equation}
  \omega ^{\pm}_{\bf k}= {k^2\over 2mS^{\pm}(k)},
  \end{equation}
 with $S^{\pm}(k)=\int d\omega S^{\pm}(k,\omega)$  the corresponding static   
structure factor.

We can establish compressibility sum rules for $S^{\pm}(k,\omega)$ by considering the 
response of the system to an external perturbation $H'_{\pm}={\lambda \over 2} 
\sum_{\bf k} (\rho_{\bf k}^{\pm}+\rho_{-\bf k}^{\pm})$ that couples
 to either the even or odd modes. We follow the analysis of Ref. \cite{huang2}
  and assume that in 
 the long wave length limit the perturbed wave function is locally the same as 
 the unperturbed one with a modulated density (even or odd for each mode). 
 We obtain 
  \begin{equation}
  \lim_{k \rightarrow0}\int d\omega {S^{\pm}(k,\omega)\over \omega}={1\over 2m v_{\pm}^2},
  \label{sr2}
  \end{equation}
with
 \begin{equation} 
  v^2_{+} = {n\over m} {\partial ^2 \epsilon_0(n,n_-)\over \partial n^2},\;\;\;\;
  v^2_{-} = {n\over m} {\partial ^2 \epsilon_0(n,n_-)\over \partial n_-^2},
   \end{equation}
   and $\epsilon_0(n,n_-)$ the ground state energy per unit volume written as a function of
   the total density $n=n_a+n_b$ and the density  
 difference $n_-=n_a-n_b$.  Since in the long wave--length limit  the sum rules are
  exhausted by
 the above quasiparticles\cite{nozieres}  [$\lim_{k\rightarrow 0} S^{\pm}(k,\omega)
 =S^{\pm}(k)\delta(\omega -
 \omega^{\pm}_{\bf k})$], the sum rules (\ref{sr1}) and (\ref{sr2}) imply that two branches 
 have energies
 $ \omega^{\pm}_{\bf k}=v_{\pm}k$, with the corresponding structure factors
given by $S^{\pm}(k)=k/2mv_{\pm}$. The low energy spectrum therefore consists of two
linear modes,  corresponding to   modulations of the density  
 with the two species ``in--phase"(even mode) and ``out--of phase"(odd mode) respectively.
  The odd mode has
 total density $n$  constant in all space, and in the spinor language
  (where the species index is treated as a spin $1/2$) corresponds to a spin 
 density wave.  We can see qualitatively that the odd modes 
 have lower frequency by perturbing around low values of the inter--species 
 interaction. For $U^{x}=0$ the even and odd modes are degenerate. 
 If we turn
 on  $U^{x}$ the odd mode will have lower expectation value $\langle U^{x} \rangle$
 since locally $\langle \rho_a({\bf x})\rho_b({\bf x}) \rangle_+
 >\langle \rho_a({\bf x})\rho_b({\bf x}) \rangle_-$.
 This is reasonable as long as the repulsive component of $U^{x}$ is  
 dominant. If we now  increase $U^{x}$ we will reach the instability
 discussed above: when $U^{x}=U^{a}$ the ground state is
 multiply degenerate with all the polarized wave functions 
 $|\Psi_0[(N-M)/2,(N+M)/2]\rangle$ ($M=-N,\cdots,N$) having the same energy.
This implies that $v_{-}\rightarrow 0$, since $ \epsilon_0(n,n_-)=\epsilon_0(n,0)$.
 Therefore the instability towards a polarized state is signaled by a vanishing velocity
 of the odd modes, or, equivalently, by a divergence of the ``compressibility'' $\kappa_- \propto
 [\partial ^2 \epsilon_0(n,n_-)/ \partial n_-^2]^{-1}$ associated with changes 
 in species polarization at constant particle number.
 
 Finally, we note that our proof is not valid in the presence of a field coupling the 
 two species of the form ${\cal R}\int d^3x\, (\psi_a^{\dagger}({\bf x}) \psi_b({\bf x}) + 
 {\rm h.c.})$. This term corresponds (in the spinor language) to a magnetic field tilted
 with respect to the direction of the otherwise fully polarized system.  If the system
 is prepared in the unstable regime with fixed $N_a$ and $N_b$ and no mechanism of interconversion
 is allowed, our proof indicates that the system will separate into two phases. This kind
 of phase separation was discussed within the Bogoliubov model by Nepomnyaschii\cite{nepo}.
  An
  infinitesimal coupling with a heat bath that allows the system to equilibrate
will equate the chemical potentials of both species and 
 induce a transition to a state of one species-only. The precise kinetics 
 and the time scale to reach equilibrium are beyond the scope of our paper.

 When ${\cal R}\neq 0$ the solutions will not be completely polarized even in the
 case where $U^{x}>U^{a}$, but that the ground state will be ``rotated" with respect
 to the quantization axis in species space. The detailed mean 
 field analysis of this problem is the subject of a forthcoming paper\cite{search}.

 We thank Chris Search, Paul Berman and Paul Julienne for very valuable discussions.

\end{document}